\newcommand{\ie}{{i.e.}}
\newcommand{\eg}{{e.g.}}

\newcommand*{\Scale}[2][4]{\scalebox{#1}{$#2$}}

\documentclass[format=sigconf, review=False]{acmart}
\usepackage{multirow}
\usepackage{algorithm}  
\usepackage{algorithmic} 

\AtBeginDocument{%
  \providecommand\BibTeX{{%
    \normalfont B\kern-0.5em{\scshape i\kern-0.25em b}\kern-0.8em\TeX}}}

\begin{document}

\title{Learning Decomposed Representation for Counterfactual Inference}

\author{Anpeng Wu, Kun Kuang, Junkun Yuan, Qiang Zhu, Yueting Zhuang, Fei Wu}
\affiliation{%
 \institution{Zhejiang University}
 \country{Hangzhou, China}}
\email{(anpwu;kunkuang;yuanjk;zhuq;yzhuang;wufei)@cs.zju.edu.cn}

\author{Bo Li}
\affiliation{%
 \institution{Tsinghua University}
 \country{China}}
\email{libo@sem.tsinghua.edu.cn}

\author{Runze Wu}
\affiliation{%
 \institution{NetEase Inc.}
 \country{Hangzhou, China}}
\email{wurunze1@corp.netease.com}

\renewcommand{\shortauthors}{Anonymous Author(s)}

\begin{abstract}
    The fundamental problem in treatment effect estimation from observational data is confounder identification and balancing. Most of the previous methods realized confounder balancing by treating all observed pre-treatment variables as confounders, ignoring further identifying confounders and non-confounders. In general, not all the observed pre-treatment variables are confounders that refer to the common causes of the treatment and the outcome, some variables only contribute to the treatment and some only contribute to the outcome. Balancing those non-confounders, including instrumental variables and adjustment variables, would generate additional bias for treatment effect estimation. By modeling the different causal relations among observed pre-treatment variables, treatment and outcome, we propose a synergistic learning framework to 1) identify confounders by learning decomposed representations of both confounders and non-confounders, 2) balance confounder with sample re-weighting technique, and simultaneously 3) estimate the treatment effect in observational studies via counterfactual inference. Empirical results on synthetic and real-world datasets demonstrate that the proposed method can precisely decompose confounders and achieve a more precise estimation of treatment effect than baselines.
    
\end{abstract}



\maketitle

\section{Introduction}
Causal inference is a powerful statistic modeling tool for explanatory analysis and plays an essential role in the decision-making process \cite{pearl2009causal}. One fundamental problem in causal inference is treatment effect estimation. For example, in the medical scenario, accurately assessing a particular drug's treatment effect on each patient will help doctors decide which medical procedure (e.g., taking the drug or not) will benefit a specific patient most. The gold standard approach for treatment effect estimation is to perform Randomized Controlled Trials (RCTs), where different treatments (i.e., medical procedures) are randomly assigned to units (i.e., patients). However, fully RCTs are often expensive \cite{kohavi2011unexpected}, unethical or even infeasible \cite{bottou2013counterfactual}. 
Hence, it is incredibly imperative and highly demanding to develop automatic statistical approaches to infer treatment effect in observational studies.


    

\begin{figure}[t]

    \centering
    \includegraphics[scale=0.64]{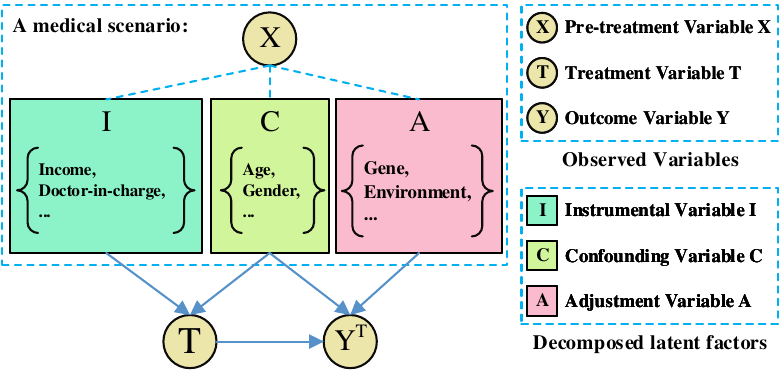}
    \caption{The intuitive illustration of our proposed causal framework w.r.t a medical scenario. Here, the historical data includes patients' pre-treatment variables $X$, the treatment $T$ and the final outcome $Y$. Among these historical data, age and gender would simultaneously affect the treatment (doctors' decision) and the outcome (patients' physical differences), hence belonging to the set of confounding factor $C$; while the income and doctor-in-charge would only affect the treatment, hence belonging to the set of instrumental factor $I$; gene and environment belong to the set of adjustment factor $A$, since they would only affect the outcome. Our proposed algoritm intends to decomepose the pre-treatment variables $X$ into the three kinds of latent factor $\{I, C, A\}$ for confounder identification and balancing. }
    \label{fig:causal}
    \vspace{-10pt}
\end{figure}

In observational studies, we denote the causal framework among the observed pre-treatment variables $X$, the treatment $T$ and the outcome $Y$, shown in Figure \ref{fig:causal}. Without loss of generality, we assume that the pre-treatment variables $X$ can be decomposed into three kinds of latent factors $\{I, C, A\}$ under an unknown joint distribution $Pr(X)=Pr(I,C,A)$, where instrumental factor $I$ only causes the treatment, confounding factor $C$ is the common cause of the treatment and the outcome, and adjustment factor $A$ only determines the outcome.
Taking the medical scenario as an example, we might collect lots of historical data from each patient, including the treatment $T$ (taking a particular drug or not), the outcome $Y$ (state of health) and patient's features $X$ (\eg, age, gender, income, gene). Among the patient's features, age and gender would simultaneously affect the treatment (doctor would consider the patient's age and gender when choosing the treatment) and the outcome (patient's age and gender would also affect his/her recovery rate), hence belonging to the set of confounding factor $C$; while the income and doctor-in-charge would only affect the treatment, but have no effect on the outcome, hence belonging to the set of instrumental factor $I$; gene and environment belong to the set of adjustment factor $A$, since they would only affect the outcome but have no effect on the treatment.

Different from RCTs, the treatment $T$ in the observational studies is not randomly assigned; instead depends on some or all attributes of unit $X$ (i.e., the factors $I$ and $C$ in Figure \ref{fig:causal}). This change could result in confounding bias: $Pr(T|X)\neq Pr(T)$. To eliminate the bias, previous methods, such as propensity score-based methods \cite{austin2011introduction,bang2005doubly,rosenbaum1983central} and variables balancing methods \cite{athey2018approximate,zubizarreta2015stable}, simply treated all observed pre-treatment variables as confounding factor (confounders) for balancing. However, back-door criteria \cite{pearl2009causality, pearl2012class} demonstrated that controlling the confounding factor is sufficient for removing that bias. In contrast, the instrumental factor's control invariably causes an increase in confounding bias if it exists. Moreover, \cite{kuang2017treatment} demonstrated that separating confounding factor and adjustment factor would reduce the estimated treatment effect variance. 
Overall, balancing the variables that mixed with non-confounders (i.e., instrumental and adjustment factors in Figure \ref{fig:causal}) would increase the bias and variance of treatment effect estimation \cite{myers2011effects, vanderweele2019principles, kuang2019treatment}.
Hence, it is indispensable to decompose the three factors for reducing the bias and variance of treatment effect estimation.

Recently, \cite{kuang2017treatment} proposed a data-driven variable decomposition method to separate adjustment variables from all observed pre-treatment variables and achieved lower variance on treatment effect estimation. Nevertheless, it ignored the decomposition of instrumental factor, which led to entanglement between instrumental and confounding factors. Moreover, it only focused on the settings with linear assumptions. \cite{hassanpour2020learning} proposed to roughly separate the pre-treatment variables into three sets $\{I,C,A\}$ with a disentangled representations learning framework (like Figure \ref{fig:causal}). However, it could not guarantee the separation between the instrumental and the confounding factors (discussed in detail in the following section), leading to the entanglement among those three factors.
Hence, precisely decomposing the instrumental, confounding and adjustment factors for confounder balancing and treatment effect estimation is still an open problem in observational studies.

To address this problem, we propose the following preliminary propositions for decomposing latent factors $\{I,C,A\}$ from pre-treatment variables $X$ as shown in Figure \ref{fig:causal}: (i) \textbf{Decomposing $A$ from $X$}: (i.a) the adjustment factor $A$ should be independent of the treatment variable $T$, i.e., $A \perp T$; and (i.b) $A$ should predict $Y$ as precisely as possible. Condition (i.a) constraints other factors not be embedded into $A$, while (i.b) restrains $A$ not be embedded into other factors. (ii) \textbf{Decomposing $I$ from $X$}: (ii.a) if the confounding factor $C$ is well balanced, one can break the dependency between $C$ and $T$, and achieve the independence between instrumental factor $I$ and outcome variable $Y$ conditional on the treatment variable $T$, i.e., $I \perp Y \mid T$; and (ii.b) $I$ should also predict $T$ as accurately as possible. Condition (ii.a) constraints other factors not be embedded into $I$, while (ii.b) restrains $I$ not be embedded into other factors. (iii) \textbf{Predicting factual and counterfactual outcomes $\{y_i^{t_i}, y_i^{1-t_i}\}$}: the decomposed representations of confounding factor $C$ and adjustment factor $A$ help to predict both factual $y_i^{t_i}$ and counterfactual outcome $y_i^{1-t_i}$.

Guided by these preliminary propositions, we further propose a synergistic learning algorithm, named Decomposed Representations for CounterFactual Regression (DeR-CFR), to jointly 1) learn and decompose the representations of the three latent factors for feature decomposition, 2) optimize sample weights for confounder balancing, and 3) learn a counterfactual regression model to predict the counterfactual outcome for treatment effect estimation in observational studies.
Our DeR-CFR algorithm is based on the standard assumptions \cite{imbens2015causal} for treatment effect estimation in observational studies, including stable unit treatment value assumption (SUTVA), unconfoundedness assumption, and overlap assumption. The main contributions in this paper are as follows:
\begin{itemize}
  \item We study the problems of confounder identification and balancing for counterfactual prediction, which is critical for accurate treatment effect estimation in observational studies.
  \item We propose a novel DeR-CFR algorithm to jointly decompose instrumental, confounding, and adjustment factors accurately, and learn counterfactual regression to estimate treatment effect in observational studies.
  \item We empirically demonstrate that our algorithm can precisely decompose the latent factors, and the results show our approach achieves a better performance of treatment effect estimation in observational studies with both synthetic and real-world datasets.
\end{itemize}

\section{Related Work}
\label{tab:related_work}
To address the confounding bias in observational studies, most of the previous methods either employ propensity score, including matching, stratification, weighting, and doubly robust \cite{rosenbaum1987model,rosenbaum1983central,li2016matching,liuyiyao2020surcey}; or directly optimize sample weight, including entropy balancing, residual balancing and stable balancing \cite{athey2018approximate,zubizarreta2015stable}. Those existing methods focus on confounder balancing alone, while ignoring the importance of confounder identification. Recently,\cite{kuang2019treatment,vanderweele2019principles} pointed out the necessity of confounder identification and selection for causal inference, due to the fact that the control of some non-confounders (\eg, variables related to the instrumental factor) would generate additional bias and amplify the variance. 
Besides, many methods \cite{pearl2009causality,vanderweele2011new} have been proposed for confounder selection, but most assume the causal structure is known prior.

\cite{johansson2016learning,shalit2017estimating} proposed a representations learning method for confounder balancing by minimizing the distribution difference between different treatment arms in embedding space. Based on these works, \cite{hassanpour2019counterfactual} proposed to optimize a context-aware importance sampling weight with representations learning jointly. Rather than taking the state-of-the-art ITE estimators to balance distribution globally, \cite{yao2018representation} proposed a local similarity preserving approach for representations learning. In this paper, we propose a decomposed representations learning approach for confounder identification along with a model-free weight schema for confounder balancing.

Our work is related to \cite{kuang2017treatment} and \cite{hassanpour2020learning}.
\cite{kuang2017treatment} proposed a data-driven variables decomposition algorithm to automatically separate confounder and adjustment factors for treatment effect estimation under a linear setting. The main limitation is that they ignored the differentiation between instrumental and confounder factors, leading to imprecise confounder identification and failing to provide an estimation of ITE. Aiming at disentangling the three latent factors $\{I,C,A\}$ from the pre-treatment variables $X$, \cite{hassanpour2020learning} proposed disentangled representations for counterfactual regression. However, the algorithm cannot guarantee to clearly decompose $I $, $ C $ and $ A $. Extremely, $I(X)^{\star}=\emptyset, C(X)^{\star}=\{I,C,A\}, A(X)^{\star}=\emptyset$ could be a possible solution of their algorithm. They cannot guarantee accurate learning disentangled representations of the instrumental factor and confounding factor, which may introduce additional bias.
Moreover, \cite{hassanpour2020learning} relied on the correct model specification on treatment and the importance sampling weights for confounder balancing. 
Our proposed algorithm is different from these methods in two ways: (i) Confounder Identification: we propose a series of decomposition regularizers to guarantee the explicit decomposition among the instrumental, confounder, and adjustment factors; (ii) Confounder Balancing: we adopt a model-free confounder balancing method to remove the confounding bias in observational data.

\section{Notations and Propositions}

In this section, we first give the notations and assumptions for treatment effect estimation in observational data, then propose a series of propositions to decompose instrumental, confounding and adjustment factors with representation leaning for treatment effect estimation.

\subsection{Notations and Assumptions}
In this paper, we focus on treatment effect estimation from observational data $\mathcal{D} = \{x_i, t_i, y_i^{t_i}\}_{i=1}^{n}$, where $n$ refers to the number of units. For each unit (e.g., patient) indexed by $i$, we observe its context characteristics $x_i\in \mathcal{X}$, its choice on treatment $t_i \in \mathcal{T}$ from a set of treatment options (e.g., \{0:placebo, 1:drug\}), and the corresponding outcome (e.g., recovery or not) $y_i^{t_i} \in \mathcal{Y}$ as a result of choosing treatment $t_i$. 

In our context, we first focus on the case of the binary treatment (for the continuous treatments, we will discuss in Section 5), and estimating the Individual Treatment Effect (ITE) of each unit $i$: 
\begin{eqnarray}
    ITE_i = y_i^1-y_i^0
\end{eqnarray}
With ITE of each unit, one can easily estimate the Average Treatment Effect (ATE) as:
\begin{eqnarray}
    \Scale[1.0]{ATE = \mathbb{E}[y^1-y^0] = \frac{1}{n}\sum_{i=1}^n ITE_i}
\end{eqnarray}
From the definition of ITE and ATE, there are two potential outcomes $y_i^0$ and $y_i^1$ for each unit $i$, however, dataset $\mathcal{D}$ only contains the observed outcome $y_i^{t_i}$ that corresponds to the treatment $t_i$, and the outcome of the alternative treatment (a.k.a. counterfactual outcome: $y_i^{1-t_i}$) is missing. This is treated as the counterfactual problem of treatment effect estimation with observational data. To address this problem, we propose a counterfactual inference framework for predicting the counterfactual outcome.

Our analysis in this paper relies on the following standard assumptions \cite{imbens2015causal} for treatment effect estimations.

{\textbf{Assumption 1: Stable Unit Treatment Value.}} The distribution of the potential outcome of one unit is assumed to be independent of the treatment assignment of another unit.

{\textbf{Assumption 2: Unconfoundedness.}} The distribution of treatment is independent of the potential outcome when given the pre-treatment variables. Formally, $T \bot \big(Y^0,Y^1\big)|X$.

{\textbf{Assumption 3: Overlap.}} Every unit should have a nonzero probability to receive either treatment status. Formally, $0 < p(T=1|X) < 1$.

\subsection{Preliminary Propositions}

As shown in Figure \ref{fig:causal}, we assume that any dataset of the form $\{X,T,Y\}$ is generated from three latent factors $\{I,C,A\}$. Inspired by the causal framework, we further generate the following preliminary propositions to support decomposition and representations learning of these three latent factors.

 \textbf{Proposition 1:} The adjustment factor would be independent of the treatment variable. Formally, $A \perp T$.

 \textbf{Proposition 2:} Under the unconfounderness assumption, controlling confounding factor can help to break the relationship between the confounding factor and the treatment variable. That is, $C \perp T$.

 \textbf{Proposition 3:} By controlling the confounding factor, the instrumental factor would become independent of the outcome, given the treatment variable. That is, if $C \perp T$, we have $I \perp Y \mid T$.

Proposition 1 can be easily understood by the definition of adjustment factor. We can denote the path between adjustment factor and treatment variable as the collider structure at $Y$: $A \rightarrow Y \leftarrow T$, hence $A \perp T$. Proposition 2 can be guaranteed by the back-door criterion \cite{pearl2009causality}. By controlling the confounder, the path between instrumental factor and outcome can be denoted as $I \rightarrow T \rightarrow Y$, hence $I \perp Y \mid T$ in proposition 3.

\textbf{Decomposing A:} Proposition 1 can only constrain that the information of other factors (i.e., $I$ and $C$) would not be embedded into $A$, but $A$ might be embedded into other factors, resulting in information leaking of $A$. To address this problem, we propose to simultaneously maximize the predictive power of $A$ on outcome $Y$ to precisely decompose the adjustment factor $A$.

\textbf{Decomposing I:} Similarly, proposition 3 can only constrain that other factors (\ie, $C$ and $A$) would not be embedded into $I$, but cannot guarantee that the information of $I$ would not be represented into other factors. In our context, we propose to jointly maximize the predictive power of $I$ on treatment $T$ for the precise decomposition of instrumental factor $I$.

By decomposing $I$ and $A$ from $X$, we can identify confounder $C$. Then, with the decomposed $C$ and $A$, we can accurately estimate the treatment effect via potential outcomes regression.

\section{DeR-CFR Algorithm}

Guided by the above preliminary propositions and analyses, we propose a novel model, named Decomposed Representations for CounterFactual Regression (DeR-CFR), to learn the decomposed representations of instrumental, confounding, and adjustment factors for confounder identification and balancing, and simultaneously learn a counterfactual regression model for treatment effect estimation. The overall architecture of our model consists of the following components:
\begin{itemize}
  \item \textbf{Three decomposed representation networks} for learning latent factors, one for each underlying factor: $I(X)$, $C(X)$ and $A(X)$.
  \item \textbf{Three decomposition and balancing regularizers} for confounder identification and balancing: the first is for decomposing $A$ from $X$ with considering $A(X) \perp T$ and $A(X)$ should predict $Y$ as precisely as possible; the second is for decomposing $I$ from $X$ via constraining $I(X) \perp Y \mid T$, and $I(X)$ should be predictive to $T$; the last is designed for simultaneously balancing confounder $C(X)$ in different treatment arms.
  \item \textbf{Two regression networks} for potential outcome prediction, one for each treatment arm: ${h^{0}}(C(X),A(X))$ and ${h^{1}}(C(X),$ \\ $A(X))$.
\end{itemize}

Our model's core components are the decomposition and balancing regularizers, which help the representation networks learn the decomposed representations of $I$, $C$, and $A$ for confounder identification, and also to improve the precision of regression networks via accurate confounder balancing with identified $C$. The decomposition and balancing regularizers are the keys to bridge the representation networks and regression networks for ITE estimation with observational data.


Next, we will describe each component of our DeR-CFR algorithm in detail.

\subsection{Decomposing $A$}

From the preliminary proposition, we know the adjustment factor should be independent of the treatment variable, $A(X) \perp T$. Considering the treatment is binary, we propose to learn the decomposed representation of adjustment factor $A(X)$ by constraining the discrepancy of its distribution between treatment arms $T=1$ and $T=0$. 
Moreover, to prevent the information of adjustment factor from being embedded into other factors, we adopt a regression model $g_A$ to maximize the predictive power of $A(X)$ on $Y$.
Here, we use $\mathcal{L}_{A}$ to denote the loss of decomposing adjustment factor as:
\begin{eqnarray}
    \Scale[1.0]{\mathcal{L}_{A} = disc(\{A(x_{i})\}_{i:t_{i}=0},\{A(x_{i})\}_{i:{t_{i}=1}})+\sum_i l[y_i, g_A(A(x_i))]}
\end{eqnarray}
where $l[y_i, g_A(A(x_i))]$ would be an $l_2$-loss for continuous outcomes and a log-loss for binary outcomes. $\left\{A\left(x_{i}\right)\right\}_{i:t_{i}=k}$ denotes the distribution of adjustment factor representation $A(X)$ with respect to the treatment arm $t=k$. Function $disc(\cdot)$ denotes the discrepancy of adjustment factor distribution between different treatment arms. Many integral probability metrics (IPMs) \cite{muller1997integral,sriperumbudur2009integral}, such as Maximum Mean Discrepancy (MMD) \cite{gretton2012kernel} and Wasserstein distance \cite{arjovsky2017wasserstein}, can be used to measure the discrepancy of distributions. In this paper, we use the MMD to calculate $disc(\cdot)$.

By minimizing this term, our model can ensure the information of the instrumental factor $I$ and the confounding factor $C$ would not be embedded into $A(X)$, since $I$ and $C$ are associated with the treatment variable. Moreover, vice versa with maximizing the predictive power of $A(X)$ on $Y$, we can ensure all the information of adjustment factor would embed to $A(X)$, hence would not be embedded into other factors.
Hence, the regularizer can help to decompose the adjustment factor.

\subsection{Decomposing $I$ and Balancing $C$}
\label{section:DecomposeIV}

From preliminary propositions, we know that if one can control the confounding factor, the instrumental factor would be independent of the outcome variable conditional on the treatment variable.

Firstly, we introduce the loss function of confounder balancing in our model. Most previous work \cite{austin2011introduction,hassanpour2019counterfactual,rosenbaum1983central} achieved confounder balancing by learning propensity score and their performance relied on the correctness of the specified propensity score model. Here, we propose to adopt a model-free method for confounder balancing.
The purpose of confounder balancing is to break the link from the confounding factor $C$ to the treatment variable $T$, that is, to make $C(X)$ become independent of $T$.
Assuming that we have the decomposed representation of confounding factor $C(X)$, we propose to achieve confounder balancing\footnote{Recently, \cite{johansson2019support,zhang2020learning} proposed alternatives for IPM (\eg, counterfactual variance) as a measure of imbalance, arguing that distributional distances are unnecessarily substantial. Therefore, there is still room for further improvement on confounder balancing.} by directly learning sample weight $\omega$ with minimizing the following objective function:
\begin{eqnarray}
\label{eq:c_balancing}
    \mathcal{L}_{C\_B} = {disc}\left(\left\{\omega_{i} \cdot C\left(x_{i}\right)\right\}_{i:t_{i}=0}, \left\{\omega_{i} \cdot C\left(x_{i}\right)\right\}_{i:t_{i}=1}\right)
\end{eqnarray}
where $\left\{\omega_{i} \cdot C\left(x_{i}\right)\right\}_{i:t_{i}=0}$ refers to the weighted distribution of $C(X)$ on the samples with $t=0$.
To avoid all the sample weights to be $zero$, we constrain the sample weight $\sum_{i:t_{i}=0} \omega_{i} = \sum_{i:t_{i}=1} \omega_{i} = 1$. If $\mathcal{L}_{C\_B}$ can be minimized to be $zero$, one can achieve the independence between $C(X)$ and $T$ by sample reweighting with the learned $\omega$.

Based on the property of the sample weight $\omega$ (\ie, $C(X)\perp T|\omega$), we can decompose the instrumental factor by conditional independence $I(X)\perp Y \mid T, \omega$.
Moreover, to prevent the information of instrumental factor from being embedded into other factors, we adopt a regression model $g_I$ to maximize the predictive power of $I(X)$ on $T$.
Then, the objective function, denoted as $\mathcal{L}_{I}$ for decomposing instrumental factor is:
\begin{eqnarray}
\nonumber    \mathcal{L}_{I} \!\!\!\!\!\! &=& \!\!\!\!\!\! \Scale[1.0]{\sum_{k=\{0,1\}} \! disc\left(\left\{\omega_{i}\! \cdot \!  I\left(x_{i}\right)\right\}_{i:y_{i}=0},\left\{\omega_{i}\! \cdot \! I\left(x_{i}\right)\right\}_{i:y_{i}=1}\right)_{i:t_{i}=k}}\\
    &+& \!\!\!\!\!\! \Scale[1.0]{\sum_i l[t_i, g_I(I(x_i))]}
\end{eqnarray}
where ${disc(\left\{\omega_{i}\cdot I\left(x_{i}\right)\right\}_{i:y_{i}=0},\left\{\omega_{i}\cdot I\left(x_{i}\right)\right\}_{i:y_{i}=1})_{i:t_{i}=k}}$ constrains the learned representation of instrumental factor $I$ to be independent of the outcome $Y$ given the treatment arm $t=k$ and sample weight $\omega$. Here, we assume the outcome variable is binary, \ie, $y_i\in \{0,1\}$. For continuous outcome, we will discuss in Section 5. 

By minimizing the term $\mathcal{L}_{I}$, our model can ensure the information of confounding factor $C$ and adjustment factor $A$ would not be embedded into $I(X)$, since $C$ and $A$ are associated with the outcome even given the treatment variable. Moreover, vice versa with maximizing the predictive power of $I(X)$ on $T$, we can ensure all instrumental factor information would be embedded into $I(X)$, hence would not be embedded into other factors. Hence, this regularizer can help to decompose the instrumental factor accurately.

\subsection{Deep Orthogonal Regularizer}
Although the representation learning based on the proposed propositions mainly contributes to the decomposition of the feature information of instrumental variables $I$, confounding variables $C$ and adjustment variables $A$, data-driven neural networks tend to overfit the training data and lead to unclean disentanglement (like DR-CFR). 
Inspired by the orthogonal regularizer in \cite{kuang2017treatment} for variable decomposition, in this paper, we employ a deep orthogonal regularizer among the three representation networks for decomposing the factors $\{I,C,A\}$. We take the representation network for instrumental factor $I$ as an example. Assuming it is with $l$ layers and let $W_{k}$ refer to the weight matrix on $k^{th}$ layer of the network. Then, we can approximate the contribution of each variable in $X$ on each dimension of representation $I(X)$ by computing $W_1\times W_2\times \cdots \times W_l$, denoted as $W_I\in \mathbb{R}^{m\times d}$, where $m$ and $d$ refer to the dimension of $X$ and $I(X)$, respectively. By averaging each row of $W_I$, we obtain $\bar{W}_I \in \mathbb{R}^m$, denoting the average contribution of each variable in $X$ on the representation $I(X)$.
Similarly, we calculate the contribution of each variable in $X$ on $C(X)$ and $A(X)$, denoted as $\bar{W}_C$ and $\bar{W}_A$.

We consider the three representation networks have the same structure. Hence, $\bar{W}_I$, $\bar{W}_C$ and $\bar{W}_A$ are the vectors that have the same dimensions. Then, we propose to achieve hard decomposition by constraining orthogonality on each pair of them. The loss is as follow:
\begin{eqnarray}
    \label{Eq:L_O}
    \mathcal{L}_{O} = \bar{W}^{T}_{I}\cdot \bar{W}_C + \bar{W}^{T}_{C}\cdot \bar{W}_A + \bar{W}^{T}_{A}\cdot \bar{W}_I
\end{eqnarray}
To guarantee the information flows of the representation networks, we softly constrain the total contribution of each $\bar{W}_I$, $\bar{W}_C$ and $\bar{W}_A$ to approximately 1, that can be found in tht regularization term $Reg$ (Section 4.5). 
The orthogonal regularizer ensures each variable's information in $X$ is either discarded or can only flow into one representation network for a hard decomposition. It can also reduce the influence of irrelevant variables on the prediction and prevent each representation network from overfitting.

\subsection{Outcome Regression}

With the decomposed representations, we propose to learn the outcome regression model for estimating the treatment effect.
Similar to \cite{johansson2016learning,shalit2017estimating,hassanpour2020learning}, we also train two regression networks for each treatment arm, $h^0$ and $h^1$, based on the observed outcomes of samples with $t_i=0$ and $t_i=1$, respectively. As guided by the graphical model in Figure \ref{fig:causal}, we train these regression models only based on the decomposed representations of $C(X)$ and $A(X)$.
\begin{eqnarray}
    \label{eq:regression}
    \mathcal{L}_R = \sum_{i} \omega_{i} \cdot l\left[y_{i}, h^{t_{i}}\left(C\left(x_{i}\right), A\left(x_{i}\right)\right)\right]
\end{eqnarray}
where the sample weight $\omega$ is learned from confounder balancing with Eq. \ref{eq:c_balancing}.

\subsection{Objective Function}

Therefore, we propose to minimize the following objective function in our DeR-CFR algorithm:
\begin{eqnarray}
    \mathcal{L}\! = \!\mathcal{L}_R + \alpha \cdot \mathcal{L}_{A}
    + \beta \cdot \mathcal{L}_{I} + \gamma \cdot \mathcal{L}_{C\_B}
    + \mu \cdot \mathcal{L}_{O}
    + \lambda \cdot Reg
\end{eqnarray}
where $Reg$ refers to the regularization term on the DeR-CFR parameters:
    \begin{eqnarray}
        Reg = \mathcal{R}_{\mathcal{W}} + \mathcal{R}_{C\_B} + \mathcal{R}_{O}
    \end{eqnarray}
where $\mathcal{R}_{\mathcal{W}}$ is the $l_2$ regularization on the parameters of subnetworks $\{ I, C, A, h^0, h^1, g_I, g_A\} $. $\mathcal{R}_{C\_B}$ restricts the sample weight $\omega$ not to be all $zero$. To guarantee the information flows of the representation networks, we use $\mathcal{R}_{O}$ to softly constrain the sum of each $\bar{W}_I$, $\bar{W}_C$, and $\bar{W}_A$  to approximately 1. The details of each regularization are introduced in the appendix for saving space.

We adopt an alternating training strategy to iteratively optimize the representations for confounder identification and sample weight for confounder balancing as:
\begin{eqnarray}
\mathcal{L_{-\omega}} \!\!\! &=& \!\!\! \mathcal{L}_R + \alpha \cdot \mathcal{L}_{A}
    + \beta \cdot \mathcal{L}_{I}
    + \mu \cdot \mathcal{L}_{O}
    + \lambda \cdot Reg\\
\mathcal{L_{\omega}} \!\!\! &=& \!\!\! \mathcal{L}_R
    + \gamma \cdot \mathcal{L}_{C\_B}
    + \lambda \cdot Reg
\end{eqnarray}

We minimize $\mathcal{L_{-\omega}}$ using stochastic gradient descent to update the parameters of the representation and hypothesis network, and minimize $\mathcal{L_{\omega}}$ to update $\omega$. The details of pseudo-code and hyper-parameters of our algorithm are provided in the appendix.

\section{Discussion on Continuous Scenes}

For continuous or multi-valued treatment and outcome, we can approximately achieve the three propositions by making treatment and outcome binary during the process of decomposing or utilizing the mutual information between $\{I, C, A\}$ with $T$ and $Y$ (CLUB \cite{cheng2020club}).

\subsection{Binarize the treatment and outcome}
For the case of the continuous treatments $T$ and continuous outcomes $Y$: In the prediction phase, we can use the regression network to predict the continuous treatment and outcome, but in the representation decomposition stage, we need to binarize the continuous treatment and outcome separately to approximately achieve the group division and some independence constraints. Without loss of generality, we use the median to divide the dataset based on the treatments $T$ and the outcomes $Y$: 
    \begin{equation}
        t^*_i := \left\{
            \begin{array}{lr}
             0, &  t_i < median(\{t_i\}) \\
             1, &  t_i \ge median(\{t_i\})
             \end{array}
            \right. 
    \end{equation}
    \begin{equation}
        y^*_i := \left\{
            \begin{array}{lr}
             0, &  y_i < median(\{y_i\}) \\
             1, &  y_i \ge median(\{y_i\})
             \end{array}
            \right. 
    \end{equation}
where $median(\{t_i\})$ refers to the median of factual outcome $\{t_i\}$ and $median(\{y_i\})$ refers to the median of factual outcome $\{y_i\}$. Similarly, other conditional division methods are also applicable. 

While decomposing $A$ and $I$ and balancing $C$, we can use $t^*_i$ and $y^*_i$ to achieve the group division and minimize the discrepancy of $I$, $C$ and $A$ in different groups by the function $disc(\cdot)$. When the treatment is continuous, we will take $C(X)$, $A(X)$ and $T$ to regress $Y$ to replace Eq. \ref{eq:regression} as follows: 
\begin{eqnarray}
    \Scale[1.0]{\mathcal{L}_R = \sum_{i} \omega_{i} \cdot l\left[y_{i}, h\left(C\left(x_{i}\right), A\left(x_{i}\right), t_{i}\right)\right]}
\end{eqnarray}

Then we can use the objective function and training strategy (Section 4.5) to optimize the representation for confounder identification and conducting counterfactual inference. In this context, we binarize the continuous treatment and outcome to extend DeR-CFR on continuous scenes.

\subsection{Utilize the mutual information}
On the case of continuous treatment and outcome, the proposition 1 can be implemented by minimizing the mutual information between $A(X)$ and $T$  (CLUB \cite{cheng2020club}): 
\begin{eqnarray}
    \Scale[0.9]{\mathcal{L}_{A} = \sum_i l[y_i, g_A(A(x_i))]} + MI\left(A\left(x\right), t\right)
\end{eqnarray}
where $MI(a, b)$ refers to the mutual information of distribution $a$ and $b$.

Besides, we can approximately achieve the conditional independence $I(X)\perp Y \mid T, \omega$ by minimizing the mutual information between $I(X)$ and $Y$ (CLUB \cite{cheng2020club}): 
\begin{eqnarray}
    \Scale[0.92] {\mathcal{L}_{I} \! = \! \sum_i l[t_i, g_I(I(x_i))] + \! \sum_{k=\{0,1\}} \! MI\left(I\left(x_{i}\right), y_i\right)_{i:t^*_{i}=k}}
\end{eqnarray}

Based on mutual information, DeR-CFR can be applied to continuous outcome scenarios but not applicable to continuous treatment (it can not balance $C$ or decompose $I$ directly). 

\section{Experiments}

\begin{table*}[htbp]
      \centering
      \caption{The results (mean$\pm $std) of treatment effect estimation on real-world data.}
      \scalebox{0.8}{
        \begin{tabular}{c|cc|cc|cc}
          \hline
          \multicolumn{7}{c}{\textbf{Within-sample}} \\
          \hline
          \textbf{Datasets} & \multicolumn{2}{c|}{\textbf{IHDP(Mean $\pm $ Std)}} & \multicolumn{2}{c|}{\textbf{Jobs(Mean $\pm $ Std)}} & \multicolumn{2}{c}{\textbf{Twins-28(Mean $\pm $ Std)}} \\
          \hline
          \hline
          Methods & PEHE & $\epsilon_{\mathrm{ATE}}$ & $\mathcal{R}_{p o l}(\pi)$ & $\epsilon_{\mathrm{ATT}}$ & PEHE & $\epsilon_{\mathrm{ATE}}$ \\
          \hline
          CFR-MMD & 0.702 $\pm $ 0.037 & 0.284 $\pm $ 0.036 & 0.194 $\pm $ 0.004 & \textbf{0.041 $\pm $ 0.015} & 0.279 $\pm $ 0.001 & 0.010 $\pm $ 0.004 \\
          CFR-WASS & 0.702 $\pm $ 0.034 & 0.306 $\pm $ 0.040 & 0.194 $\pm $ 0.004 & {0.041 $\pm $ 0.016} & 0.277 $\pm $ 0.001 & 0.021 $\pm $ 0.001 \\
          CFR-ISW & 0.598 $\pm $ 0.028 & 0.210 $\pm $ 0.028 & 0.189 $\pm $ 0.006 & 0.041 $\pm $ 0.017 & 0.279 $\pm $ 0.001 & 0.036 $\pm $ 0.002 \\
          SITE & 0.609 $\pm $ 0.061 & 0.259 $\pm $ 0.091 & 0.224 $\pm $ 0.005 & 0.064 $\pm $ 0.022 & 0.279 $\pm $ 0.001 & {0.037 $\pm $ 0.003} \\
          DR-CFR & 0.657 $\pm $ 0.028 & 0.240 $\pm $ 0.032 & {0.199 $\pm $ 0.006}  & 0.064 $\pm $ 0.026 & 0.276 $\pm $ 0.001 & \textbf{0.006 $\pm $ 0.002} \\
          DeR-CFR & \textbf{0.444 $\pm $ 0.020} & \textbf{0.130 $\pm $ 0.020} & \textbf{0.187 $\pm $ 0.037} & {0.053 $\pm $ 0.084} & \textbf{0.276 $\pm $ 0.001} & 0.008 $\pm $ 0.003 \\
          \hline
          \hline
          \multicolumn{7}{c}{\textbf{Out-of-sample}} \\
          \hline
          \textbf{Datasets} & \multicolumn{2}{c|}{\textbf{IHDP(Mean $\pm $ Std)}} & \multicolumn{2}{c|}{\textbf{Jobs(Mean $\pm $ Std)}} & \multicolumn{2}{c}{\textbf{Twins-28(Mean $\pm $ Std)}} \\
          \hline
          \hline
          Methods & PEHE & $\epsilon_{\mathrm{ATE}}$ & $\mathcal{R}_{p o l}(\pi)$ & $\epsilon_{\mathrm{ATT}}$ & PEHE & $\epsilon_{\mathrm{ATE}}$ \\
          \hline
          CFR-MMD & 0.795 $\pm $ 0.078 & 0.309 $\pm $ 0.039 & {0.222 $\pm $ 0.019} & {0.084 $\pm $ 0.028} & 0.284 $\pm $ 0.005 & 0.010 $\pm $ 0.004 \\
          CFR-WASS & 0.798 $\pm $ 0.088 & 0.325 $\pm $ 0.045 & 0.225 $\pm $ 0.023 & 0.102 $\pm $ 0.047 & 0.281 $\pm $ 0.005 & 0.023 $\pm $ 0.003 \\
          CFR-ISW & 0.715 $\pm $ 0.102 & 0.218 $\pm $ 0.031 & 0.225 $\pm $ 0.024 & 0.089 $\pm $ 0.033 & 0.283 $\pm $ 0.006 & 0.039 $\pm $ 0.004 \\
          SITE & 1.335 $\pm $ 0.698 & 0.341 $\pm $ 0.116 & 0.229 $\pm $ 0.023 & \textbf{0.074 $\pm $ 0.028} & 0.283 $\pm $ 0.006 & {0.040 $\pm $ 0.004} \\
          DR-CFR & 0.789 $\pm $ 0.091 & 0.261 $\pm $ 0.036 & 0.235 $\pm $ 0.015 & 0.119 $\pm $ 0.045 & 0.280 $\pm $ 0.005 & {0.009 $\pm $ 0.003} \\
          DeR-CFR & \textbf{0.529 $\pm $ 0.068} & \textbf{0.147 $\pm $ 0.022} & \textbf{0.208 $\pm $ 0.062} & 0.093 $\pm $ 0.032 & \textbf{0.279 $\pm $ 0.005} & \textbf{0.008 $\pm $ 0.004} \\
          \hline
          \end{tabular}%
      }
      \label{tab:real}%
    \end{table*}%

\begin{table}[t]
  \centering
  \caption{Results (mean$\pm$std) of ablation studies on IHDP dataset (\checkmark\  refers to keeping the component in DeR-CFR).}
  \scalebox{0.8}{
    \begin{tabular}{c c c c | c c }
    \hline

    \multirow{2}[2]{*}{$\mathcal{L}_{A}$} &
    \multirow{2}[2]{*}{$\mathcal{L}_{I}$} & 
    \multirow{2}[2]{*}{$\mathcal{L}_{C\_B}$} & 
    \multirow{2}[2]{*}{$\mathcal{L}_{O}$} &

    \multicolumn{2}{c}{PEHE} \\
    \cline{5-6}          &       &       &       & Within-sample & Out-of-sample \\
    \hline
      & \checkmark & \checkmark & \checkmark & 0.635 $\pm $ 0.035 & 0.858 $\pm $ 0.133 \\
    \hline
    \checkmark &   & \checkmark & \checkmark & 0.479 $\pm $ 0.030 & 0.560 $\pm $ 0.071 \\
    \hline
    \checkmark & \checkmark &   & \checkmark & 0.482 $\pm $ 0.039 & 0.565 $\pm $ 0.075 \\
    \hline
    \checkmark & \checkmark & \checkmark &   & 0.478 $\pm $ 0.033 & 0.542 $\pm $ 0.053 \\
    \hline
    \checkmark & \checkmark & \checkmark & \checkmark & \textbf{0.444 $\pm $ 0.020} & \textbf{0.529 $\pm $ 0.068} \\
    \hline
    \end{tabular}%
    }
    \label{tab:ablation}%
\end{table}%

\begin{table*}[t]
  \centering
  \caption{Results (mean $\pm $ std) on synthetic data under different settings (Binary\_$m_I$\_$m_C$\_$m_A$\_$n$). }
  \scalebox{0.7}{
    \begin{tabular}{c|cc|cc|cc|cc}
      \hline
      \multicolumn{9}{c}{\textbf{Within-sample}} \\
      \hline
      {\textbf{Setting}} & 
      \multicolumn{2}{c|}{\textbf{Binary\_8\_8\_8\_3000}} & 
      \multicolumn{2}{c|}{\textbf{Binary\_8\_8\_8\_10000}} & 
      \multicolumn{2}{c|}{\textbf{Binary\_16\_16\_16\_3000}} & 
      \multicolumn{2}{c}{\textbf{Binary\_16\_16\_16\_10000}} \\
      \hline
      \hline
      Methods & PEHE & $\epsilon_{\mathrm{ATE}}$ &PEHE & $\epsilon_{\mathrm{ATE}}$ &PEHE & $\epsilon_{\mathrm{ATE}}$ &PEHE & $\epsilon_{\mathrm{ATE}}$ \\
      \hline
      CFR-MMD & 0.384 $\pm $ 0.004 & {0.015 $\pm $ 0.006} & 0.276 $\pm $ 0.004 & {0.008 $\pm $ 0.003} & 0.491 $\pm $ 0.005 & 0.021 $\pm $ 0.008 & 0.399 $\pm $ 0.005 & {0.012 $\pm $ 0.005} \\
      CFR-WASS & 0.378 $\pm $ 0.004 & 0.016 $\pm $ 0.006 & 0.277 $\pm $ 0.004 & {0.008 $\pm $ 0.002} & 0.513 $\pm $ 0.007 & \textbf{0.011 $\pm $ 0.005} & 0.408 $\pm $ 0.005 & 0.015 $\pm $ 0.005 \\
      CFR-ISW & 0.383 $\pm $ 0.005 & 0.035 $\pm $ 0.007 & 0.279 $\pm $ 0.004 & 0.013 $\pm $ 0.002 & 0.538 $\pm $ 0.003 & {0.014 $\pm $ 0.005} & 0.441 $\pm $ 0.005 & 0.034 $\pm $ 0.005 \\
      SITE & 0.550 $\pm $ 0.007 & 0.075 $\pm $ 0.013 & 0.497 $\pm $ 0.006 & 0.035 $\pm $ 0.012 & 0.585 $\pm $ 0.005 & 0.035 $\pm $ 0.012 & 0.608 $\pm $ 0.006 & 0.041 $\pm $ 0.014 \\
      DR-CFR & 0.377 $\pm $ 0.002 & 0.027 $\pm $ 0.008 & 0.288 $\pm $ 0.005 & 0.022 $\pm $ 0.007 & 0.544 $\pm $ 0.004 & 0.023 $\pm $ 0.010 & 0.427 $\pm $ 0.015 & 0.043 $\pm $ 0.019 \\
      DeR-CFR & \textbf{0.325 $\pm $ 0.002} & \textbf{0.014 $\pm $ 0.006} & \textbf{0.234 $\pm $ 0.003} & \textbf{0.007 $\pm $ 0.002} & \textbf{0.404 $\pm $ 0.003} & \textbf{0.011 $\pm $ 0.004} & \textbf{0.307 $\pm $ 0.002} & \textbf{0.006 $\pm $ 0.002} \\
      \hline
      \hline
      \multicolumn{9}{c}{\textbf{Out-of-sample}} \\
      \hline
      {\textbf{Setting}} & 
      \multicolumn{2}{c|}{\textbf{Binary\_8\_8\_8\_3000}} & 
      \multicolumn{2}{c|}{\textbf{Binary\_8\_8\_8\_10000}} & 
      \multicolumn{2}{c|}{\textbf{Binary\_16\_16\_16\_3000}} & 
      \multicolumn{2}{c}{\textbf{Binary\_16\_16\_16\_10000}} \\
      \hline
      \hline
      Methods & PEHE & $\epsilon_{\mathrm{ATE}}$ &PEHE & $\epsilon_{\mathrm{ATE}}$ &PEHE & $\epsilon_{\mathrm{ATE}}$ & PEHE & $\epsilon_{\mathrm{ATE}}$ \\
      \hline
      CFR-MMD & 0.465 $\pm $ 0.006 & 0.062 $\pm $ 0.021 & 0.327 $\pm $ 0.006 & 0.021 $\pm $ 0.008 & 0.574 $\pm $ 0.007 & 0.036 $\pm $ 0.012 & 0.463 $\pm $ 0.006 & \textbf{0.018 $\pm $ 0.006} \\
      CFR-WASS & 0.469 $\pm $ 0.011 & 0.063 $\pm $ 0.021 & 0.320 $\pm $ 0.006 & 0.016 $\pm $ 0.007 & 0.553 $\pm $ 0.006 & \textbf{0.028 $\pm $ 0.009} & 0.469 $\pm $ 0.005 & \textbf{0.018 $\pm $ 0.007} \\
      CFR-ISW & 0.461 $\pm $ 0.005 & 0.058 $\pm $ 0.021 & 0.334 $\pm $ 0.006 & 0.017 $\pm $ 0.007 & 0.553 $\pm $ 0.006 & {0.034 $\pm $ 0.012} & 0.501 $\pm $ 0.005 & 0.040 $\pm $ 0.007 \\
      SITE & 0.561 $\pm $ 0.005 & 0.077 $\pm $ 0.020 & 0.506 $\pm $ 0.006 & 0.021 $\pm $ 0.009 & 0.588 $\pm $ 0.007 & 0.050 $\pm $ 0.016 & 0.612 $\pm $ 0.009 & 0.049 $\pm $ 0.013 \\
      DR-CFR & 0.469 $\pm $ 0.011 & 0.063 $\pm $ 0.024 & 0.333 $\pm $ 0.006 & 0.030 $\pm $ 0.009 & 0.551 $\pm $ 0.008 & 0.037 $\pm $ 0.014 & 0.486 $\pm $ 0.011 & 0.044 $\pm $ 0.019 \\
      DeR-CFR & \textbf{0.409 $\pm $ 0.009} & \textbf{0.046 $\pm $ 0.017} & \textbf{0.286 $\pm $ 0.007} & \textbf{0.012 $\pm $ 0.006} & \textbf{0.485 $\pm $ 0.006} & \textbf{0.028 $\pm $ 0.010} & \textbf{0.376 $\pm $ 0.006} & \textbf{0.018 $\pm $ 0.005} \\ 
      \hline
      \end{tabular}%
  }
  \label{tab:Syn}%
\end{table*}%

\begin{table}[t]
  \centering
  \caption{Results (mean $\pm $ std) on continuous setting.}
  \scalebox{0.8}{
    \begin{tabular}{ c | c c }
    \hline

    \multirow{2}[2]{*}{\textbf{Methods}} & 
    \multicolumn{2}{c}{\textbf{MSE} } \\
    \cline{2-3}& Within-sample & Out-of-sample \\
    \hline
    CFR-MMD & {0.044 $\pm $ 0.008} & {0.048 $\pm $ 0.022} \\
    \hline
    CFR-WASS & {0.046 $\pm $ 0.006} & {0.054 $\pm $ 0.012} \\
    \hline
    CFR-ISW & {0.058 $\pm $ 0.014} & {0.064 $\pm $ 0.009} \\
    \hline
    SITE & {0.235 $\pm $ 0.033} & {0.216 $\pm $ 0.059} \\
    \hline
    DR-CFR & {0.098 $\pm $ 0.044} & {0.106 $\pm $ 0.032} \\
    \hline
    DeR-CFR & \textbf{0.026 $\pm $ 0.002} & \textbf{0.028 $\pm $ 0.004} \\
    \hline
    \end{tabular}%
    }
    \label{tab:cont}%
\end{table}%

\subsection{Baselines}
We compare the proposed algorithm (\textbf{DeR-CFR}) with the following baselines. (1) \textbf{CFR-MMD} and \textbf{CFR-WASS} \cite{johansson2016learning,shalit2017estimating}: CounterFactual Regression with MMD and Wasserstein metrics; (2) \textbf{CFR-ISW} \cite{hassanpour2019counterfactual}: CounterFactual Regression with Importance Sampling Weights; (3) \textbf{SITE} \cite{yao2018representation}: local Similarity
preserved Individual Treatment Effect estimator; and (4) \textbf{DR-CFR} \cite{hassanpour2020learning}: Disentangled Representations for CounterFactual Regression. For continuous scenes, we binarize the continuous treatment to run these baselines and utilize the learned representation to regress the continuous outcomes. 

\subsection{Experiments on Real Dataset}

\subsubsection{Dataset.} In order to evaluate the proposed method, we conduct the experiment on three real-world datasets that are adopted in \cite{yao2018representation}: IHDP, Jobs and Twins-28. IHDP aims to evaluate the effect of specialist home visits on premature infants' future cognitive test scores and Jobs aims to estimate the effect of job training programs on employment status.

The twins dataset is derived from the all twins born in the USA between the year of 1989 and 1991 \cite{almond2005costs}. When a unit is the heavier one in the twins, the treatment is ${{t}_{i}}=1$, and the lighter one is ${{t}_{i}}=0$. Besides, we obtained 28 variables related to parents, pregnancy, and birth. The outcome is the children’s mortality after one year. We focus on same-sex twins weighing less than 2000g and without missing features. The final dataset contains 5271 records. To develop the instrument variables, we generate 38-dimension variables for each unit: $X = \{ {X_1},{X_2},...,{X_{38}}\}$, where $X_{1}, X_{2}, \cdots, X_{10} \sim \mathcal{B}(5,0.5)$ and  $\{ {X_{11}},{X_{12}},...,{X_{38}}\}$ comes from the original data.
 The treatment assignment strategy is: $t_{i} | x_{i} \sim {\rm{Bern}} \left( {{\rm{ }}{\rm{sigmoid}}{\rm{ }}\left( {{w^T}{X_{AB}} + n} \right)} \right)$, where ${w^T} \sim {U}\left( {{{( - 0.1,0.1)}^{44 \times 1}}} \right)$ and $n \sim {N}(0,0.1)$.
We conduct our experiments on the 10 realizations of Twins with a 63/27/10 proportion of train/validation/test splits. 

\subsubsection{Metrics.} 
On IHDP and Twins, we adopt the Precision in Estimation of Heterogeneous Effect (PEHE) \cite{hill2011bayesian,hassanpour2020learning} as the individual-level performance metric, where PEHE =$\Scale[0.9]{\sqrt {\frac{1}{N}\sum_{i = 1}^N {{{\left( {(\hat y_i^1 - \hat y_i^0) - (y_i^1 - y_i^0)} \right)}^2}} } }$. For population-level, we adopt the bias of ATE prediction $\epsilon_{\mathrm{ATE}} = |ATE - \widehat{ATE}|$ to evaluate performance, where ATE = $\mathbb E(y^1) - \mathbb E(y^0)$. 

On Jobs dataset, there is no ground truth for counterfactual outcomes, so the policy risk \cite{shalit2017estimating} is adopted, which is defined as: 
$\mathcal{R}_{pol} = 1 - \mathbb E \left[ {{y^1}|\pi_f (x) = 1, t=1} \right]{\mathcal P}\left( {\pi_f (x) = 1} \right) - \mathbb E\left[ {{y^0}|\pi_f (x) = 0, t=0} \right]$ \\
${\mathcal P}\left( {\pi_f (x) = 0} \right) $, 
where $\pi_f(x)=1$ if $\hat{y}_{1}-\hat{y}_{0}>0$ and $\pi_f (x)=0$, otherwise. The policy risk measures the expected loss if the treatment is taken according to the ITE estimation. For PEHE and policy risk, the smaller value is, the better the performance. 

\subsubsection{Results.} 
We report the results, including the mean and standard deviation (std) of treatment effect over 100 replications on IHDP, 10 replications on Jobs and Twins-28 datasets in Table \ref {tab:real}. The results show that in comparison with state-of-the-art methods, DeR-CFR outperforms all baselines and achieves a significant improvement on PEHE and $\epsilon_{ATE}$ measures on the IHDP dataset. On Jobs and Twins, DeR-CFR has comparable performance to the state-of-art in estimating treatment effects. Our algorithm does not achieve such significant improvement on Jobs and Twins-28 than IHDP data; the main reason we analyzed is that (i) on Jobs, most of the manually selected variables may be confounding variables, DeR-CFR would be not prominent compared with other baseline in this case; (ii) on Twins, all variables are discrete and most units have similar data, which leads to the low improvement of our DeR-CFR algorithm. 

Table \ref{tab:ablation} investigates the effects of each module of the DeR-CFR by conducting ablation experiments on IHDP.
From Tabel \ref{tab:real} and Table \ref{tab:ablation}, we can draw the following conclusions: (i) With explicitly learning the decomposed representations, DeR-CFR achieves better performance than DR-CFR, which cannot guarantee the disentanglement of different factors. (ii) Each component in our DeR-CFR is necessary, since missing any one of them would confuse the decomposed representations learning and damage the performance of ITE estimation on IHDP dataset. 

\subsection{Experiments on Synthetic Dataset}

\begin{figure*}[t]
  \centering
  \includegraphics[scale=0.48]{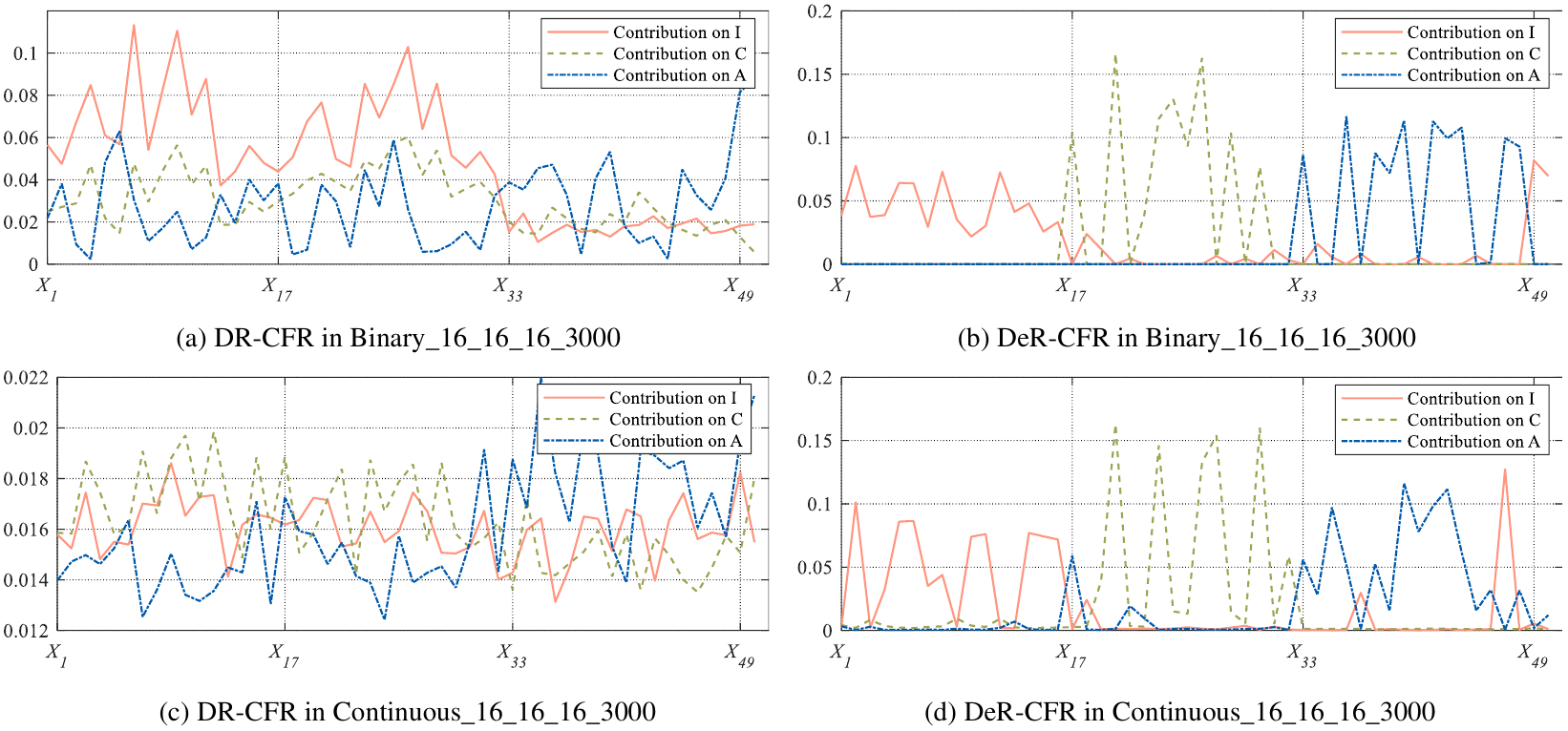}
  \vspace{-3pt}
  \caption{Visualization of the contribution of each variable in $X$ on the decomposed representations of $I$, $C$ and $A$ under the settings with Binary\_16\_16\_16\_3000 (sub-figures a,b) and Continuous\_16\_16\_16\_3000 (sub-figures c,d), where $X_I = \{X_{1}\cdots,X_{16}\}$, $X_C = \{X_{17}\cdots,X_{32}\}$ and $X_A = \{X_{33}\cdots,X_{48}\}$ are the true underlying factors of $I$, $C$ and $A$.}
  \label{fig:disentangle_single}
\end{figure*}


\begin{figure}[t]
  \centering
  \includegraphics[scale=0.20]{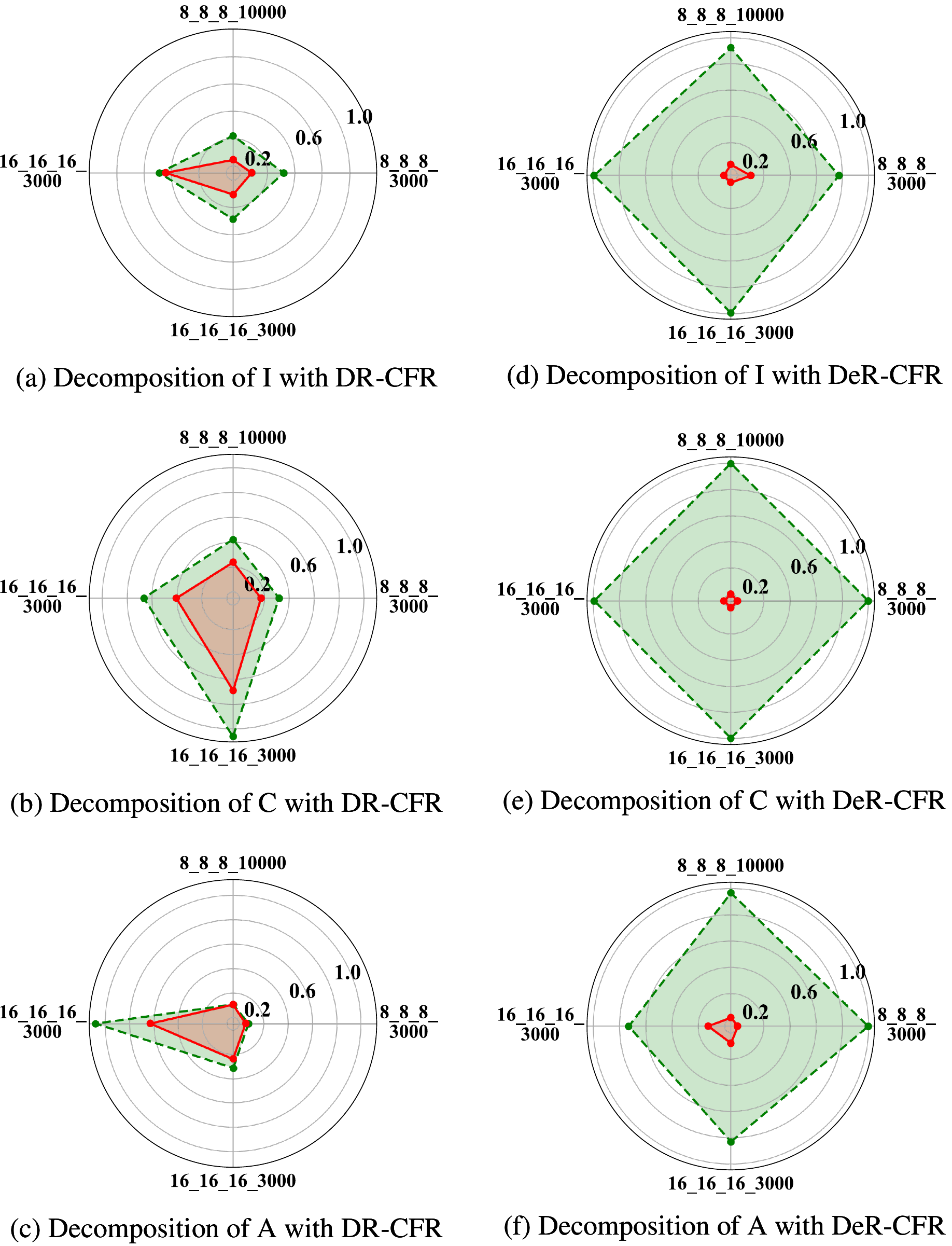}
  \caption{Radar charts that visualize the disentangled/decomposed representations of all three underlying factors from DR-CFR (sub-figures a,b,c) and DeR-CFR (sub-figures d,e,f) methods. Each vertex on the polygons denotes an experimental setting with form Binary\_$m_I$\_$m_C$\_$m_A$\_$n$. The green and red plots denote the average contribution of true variables and other variables in $X$ on the representation of each factor, respectively.}
  \label{fig:radar}
  \vspace{-20pt}
\end{figure}

\subsubsection{Dataset.} To generate synthetic datasets, we design two different sample sizes $n = {\rm{\{ }}3000,10000\} $ and two dimensional settings $\{m_I,m_C,m_A\} $=$\{ 8, 8, 8\}$ or $\{ 16, 16, 16\}$, where $m_I, m_C$, and $m_A$ denote the dimensions of instrumental variables, confounding variables and adjustment variables, respectively. Thus, the total dimension of pre-treatment variables is $m={{m}_{I}}+{{m}_{C}}+{{m}_{A}}+{{m}_{D}}$, where ${{m}_{D}}=2$ denotes two noise variables. We generate samples from independent Normal distributions $ X_{1}, X_{2}, \cdots, X_{m} \sim \mathcal{N}(0,1)$. 

\noindent \textbf{Binary Setting:} In this paper, we focus on the setting with binary treatment and binary outcome. We first generate binary treatment $t= binomial (1, 1/(1+e^{-z}))$, where $z=\frac{1}{10} \theta_{t} \times 
X_{I C}+\varepsilon$, $X_{I C}$ denotes the variables in $X$ that belongs to $I$ and $C$. Then, generate binary outcomes corresponding to different treatment arms as $y^{0}={\rm{sign}}\left(\max \left(0, z^{0}-\bar{z}^{0}\right)\right)$ and $y^{1}={\rm{sign}}\left(\max \left(0, z^{1}-\bar{z}^{1}\right)\right)$, where $z^{0}=\frac{1}{10} \frac{\theta_{y 0} \times X_{C A}}{m_{C}+m_{A}}$ and $z^{1}=\frac{1}{10} \frac{\theta_{y 1} \times X_{C A}^{2}}{m_{C}+m_{A}}$. In addition, ${\theta _t} \sim \mathcal{U}\left((8,16)^{m_I+m_C}\right) , \ \ {\theta _{y0}},{\theta _{y1}} \sim \mathcal{U}\left((8,16)^{m_C+m_A}\right), \varepsilon \sim \mathcal{N}(0,1)$. We use Binary\_$m_I$\_$m_C$\_$m_A$\_$n$ to denote different experimental settings. In each setting, we do experiments with 10 replications, and report the mean and standard deviation (std) on PEHE and $\epsilon_{\mathrm{ATE}}$.

\noindent \textbf{Continuous Setting:} Our algorithm can be also applied for continuous treatment and outcome as we discussed in Section 5. Here, we also generate continuous treatment and outcome as $t = 1/(1+e^{-z})$ and $y = z^0 + t * z^1 + \epsilon_y$, where $\epsilon_y \sim \mathcal{N}(0,0.1)$. We use Continuous\_$m_I$\_$m_C$\_$m_A$\_$n$ to denote continuous settings.


\subsubsection{Results of treatment effect estimation.}
In binary setting, we compare our DeR-CFR with the contending baselines under different settings and report the results in Table \ref{tab:Syn}. We see that DeR-CFR outperforms other state-of-the-art methods in PEHE and $\epsilon_{\mathrm{ATE}}$. Moreover, with the explicit decomposition of instrumental, confounding and adjustment factors during representations learning, the performance of DeR-CFR is much better than DR-CFR.

In continuous setting with Continuous\_16\_16\_16\_3000, we report the mean square error (MSE) of counterfactual outcome prediction (detailed definition is introduce in the appendix) with 10 independent replications in Table \ref{tab:cont}. From the result, we can conclude that considering the decomposed representation of confounders and non-confounders, our DeR-CFR can achieve the best performance than baselines on counterfactual regression.

\subsubsection{Results on Decomposed Representation.}
To evaluate the performance of decomposed representation learning, we calculate the average contribution of each variable in $X$ on the representation of each factor, \ie, $\bar{W}_I, \bar{W}_C, \bar{W}_A \in \mathbb{R}^m$ as described in the previous section. Figure \ref{fig:disentangle_single} reports the results under settings of binary\_16\_16\_16\_3000 (Figure \ref{fig:disentangle_single}(a,b)) and continuous\_16\_16\_16\_3000 (Figure \ref{fig:disentangle_single}(c,d)).
It is evident in Figure \ref{fig:disentangle_single} that our DeR-CFR algorithm can precisely identify the three underlying factors, while the baseline DR-CFR fails to disentangle those factors. This result vilifies the motivation of the proposed DeR-CFR and is consistent with our analysis on the comparison of DeR-CFR and DR-CFR algorithms in the previous section.

Similar to the setting in DR-CFR \cite{hassanpour2020learning}, we also plot the radar charts on the representation of each factor for comparison in Figure \ref{fig:radar}. For example, in Figure \ref{fig:radar}(a), we calculate the average contribution of true variables of $I$ in $X$, \ie, $X_I=\{X_1,\cdots,X_{16}\}$ on the representation of $I$ (plotted with dotted green), compared with the average contribution of other variables in $X$, \ie, $X\backslash X_I=\{X_{17},\cdots,X_{48}\}$ on the representation of $I$ (plotted with red) under different settings. From the results, we can conclude that with explicit decomposed representation, our DeR-CFR achieves much better decomposed/disentangled representations of all three underlying factors than DR-CFR. This is the key reason that our DeR-CFR can obtain significant improvement on treatment effect estimation than DR-CFR, as shown in Table \ref{tab:Syn}. 

\subsection{Hyper-parameters Analysis}
\begin{figure*}[t]

    \centering
    \includegraphics[scale=0.4]{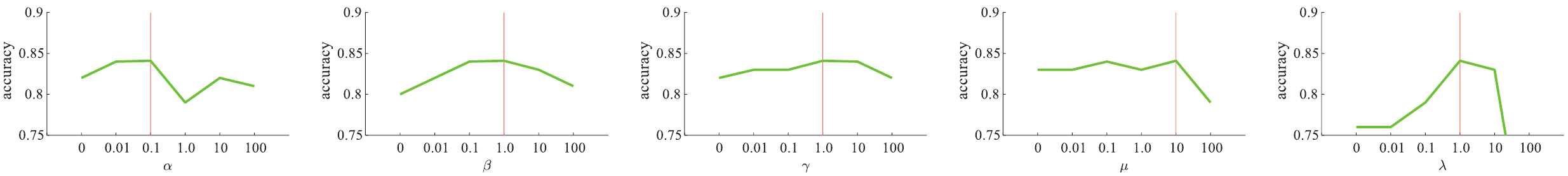}
    \caption{Hyper-parameter sensitivity analysis on $\{\alpha, \beta, \gamma, \mu, \lambda\}$. The green lines show the accuracy of the these parameters within the specified range $\{0, 0.01, 0.1, 1.0, 10, 100\}$. The red line indicates the best parameters for the setting.}
    \label{fig:alter_param}
    
\end{figure*}

Given the complex multi-term objective function (Section 4.5) in DeR-CFR, we study the impact of each item on the accuracy of the potential outcomes under setting Binary\_16\_16\_16\_3000 by changing $\{\alpha, \beta, \gamma, \mu, \lambda\}$ in the scope $\{0, 0.01, 0.1, 1.0, 10, 100\}$. The result in Figure \ref{fig:alter_param} demonstrates that the performance of DeR-CFR is mostly affected by changing in $\alpha$ and $\lambda$, reflecting the fact that decomposing adjustment factor $A$ accurately will greatly contribute to the improvement of performance and limiting the complexity of the model is necessary. $\mu$ will guarantee the decomposition of three latent factors, which not only help each representation network to select information, but will also prevent the model from overfitting. $\beta$ and $\gamma$ may not affect the accuracy obviously, but they are an essential condition for confounder identification. 
With hyper-parameters analysis, we can choose the best hyper-parameters for experiments.

\begin{table}[t]
  \centering
  \caption{Mutual Information interpretation for DeR-CFR.}
  \scalebox{0.78}{
    \begin{tabular}{c | c | c | c | c }
    
    \hline
    \multirow{2}[2]{*}{\textbf{MI}} & \multicolumn{2}{c|}{\textbf{DR-CFR}} & \multicolumn{2}{c}{\textbf{DeR-CFR}} \\
    
    \cline{2-5} & \textbf{$T$} & \textbf{$Y$} & \textbf{$T$} & \textbf{$Y$} \\
    
    \hline
    \hline
    
    \textbf{$I$} & 0.0267 $ \sim $ 0.0472 & 0.0158 $ \sim $ 0.0150 & 0.1993 $ \sim $ 0.3874 & 0.0010 $ \sim $ 0.0823 \\
    \textbf{$C$} & 0.0157 $ \sim $ 0.2115 & 0.0141 $ \sim $ 0.2004 & 0.3729 $ \sim $ 0.4561 & 0.3599 $ \sim $ 0.4439 \\
    \textbf{$A$} & 0.0001 $ \sim $ 0.0004 & 0.0001 $ \sim $ 0.0004 & 0.0439 $ \sim $ 0.2113 & 0.2494 $ \sim $ 0.4151 \\
    \textbf{$X$} & 0.4892 $ \sim $ 0.6485 & 0.3365 $ \sim $ 0.6605 & 0.4892 $ \sim $ 0.6485 & 0.3365 $ \sim $ 0.6605 \\
    
    \hline
    
    \end{tabular}%
    }
    \label{tab:MI}%
\end{table}%

\subsection{Mutual Information Interpretation.}
We also demonstrate the mutual information with lower and upper bound (CLUB \cite{cheng2020club}) under setting Binary\_16\_16\_16\_3000. The results are summarized in Table \ref{tab:MI}, which demonstrates the learned $I$ from DeR-CFR is weakly correlated with $Y$ but highly correlated with $T$, and the learned $A$ from DeR-CFR is weakly related to T but highly correlated with Y. Consistent with the results in Figure \ref{fig:disentangle_single}, the mutual information between factors $\{I, C, A\}$ with treatment $T$ and $Y$ shows DeR-CFR does decompose instrumental variables $I$, confounding variables $C$ and adjustment variables $A$. In addition, the results show that the representation network $I$ in DR-CFR overfits the training data and the learned $A$ from DR-CFR may be empty (i.e., $A = \emptyset$) without explicit decomposition constraints.

\section{Conclusion}
In this paper, we focus on the problem of estimating treatment effect in observational studies. We argue that previous methods mainly focus on confounder balancing, while ignoring the importance of confounder identification. Although some promising algorithms have been proposed for confounder separation/disentanglement, they cannot guarantee the decomposition of instrumental factor and confounding factor. Hence, we propose a Decomposed representations learning algorithm for CounterFactual Regression (DeR-CFR) with explicit decomposition constraints for confounder identification and balancing, and simultaneously estimate the treatment effect via counterfactual inference. Empirical results demonstrate the advantages of the DeR-CFR algorithm compared with state-of-the-art methods.

\bibliographystyle{ACM-Reference-Format}
\bibliography{reference}

\appendix

\section{The Regularization Term on DeR-CFR Parameters}
In the DeR-CFR Algorithm, $Reg$ refers to the regularization term on network parameters: 
    \begin{eqnarray}
        Reg = \mathcal{R}_{\mathcal{W}} + \mathcal{R}_{C\_B} + \mathcal{R}_{O}
    \end{eqnarray}
Next, we describe each component of $Reg$ in detail.

\subsection{The regularization on the network parameters.}
In the DeR-CFR Algorithm, we add $l_2$ regularization on the parameters of subnetworks $\{ I, C, A, h^0, h^1, g_I, g_A\}$ to prevent over-fitting:
    \begin{eqnarray}
        \mathcal{R}_{\mathcal{W}} = l_{2}\left( \mathcal{W}(I, C, A, h^0, h^1, g_I, g_A) \right)
    \end{eqnarray}
    
The regularization term is generally a monotonically increasing function of the model complexity. We believe that the model will have lower complexity and better robustness when the model's parameter value is small enough. To prevent overfitting, we penalize the immense value in the network parameters $\mathcal{W}(I, C, A, h^0, h^1, g_I, g_A)$ by $l_2$ regularization.

\subsection{The regularization on the sample weight.}
$\mathcal{R}_{C\_B}$ restricts the sample weight $\omega$ not to be all $zero$ and approximately 1: 
    \begin{eqnarray}
        \mathcal{R}_{C\_B} = \left( \sum_{i:t_{i}=0} \omega_{i} - 1 \right)^2 + \left( \sum_{i:t_{i}=1} \omega_{i} - 1 \right)^2 
    \end{eqnarray}
    
To avoid all the sample weights to be $zero$ and maintain original quantity allocation on each treatment arm, we constrain the sample weight $\sum_{i:t_{i}=0} \omega_{i} = \sum_{i:t_{i}=1} \omega_{i} = 1$.

\subsection{The regularization on the orthogonal regularizer.}
    
    While minimizing $\mathcal{L}_{O}$ (in Eq. \ref{Eq:L_O}), the deep orthogonal regularizer may lead to the result $\bar{W}^{k}_{I} = \bar{W}^{k}_{C} = \bar{W}^{k}_{A} = 0$ for all dimension $k$. To guarantee the information flows of the representation networks, we softly constrain the sum of each $\bar{W}_I$, $\bar{W}_C$, and $\bar{W}_A$  to approximately 1: 
    \begin{small}
        \begin{equation}
            \begin{aligned}
                \mathcal{R}_{O} \! = \left( \sum_{k=1}^{m} \bar{W}^{k}_{I} - 1 \right)^2 \! + \! \left( \sum_{k=1}^{m} \bar{W}^{k}_{C} - 1 \right)^2 
                 \! + \! \left( \sum_{k=1}^{m} \bar{W}^{k}_{A} - 1 \right)^2 
            \end{aligned}
        \end{equation}
    \end{small}
    
\section{MSE: The performance metric in continuous scenes}

When the treatment and outcome are continuous, the goal of counterfactual outcome prediction is to get a counterfactual estimation function $h\left(C\left(x_{i}\right), A\left(x_{i}\right), t_{i}\right)$ that is close to true response function $f\left(x_{i}, t_{i}\right) = z^0\left(x_{i}\right) + t_i * z^1\left(x_{i}\right)$, typically measured by Mean Square Error (MSE) \cite{hartford2017deep}:

\begin{equation}
    MSE=\frac{1}{n}{{\sum\limits_{i}{\left( h\left( C\left( {{x}_{i}} \right),A\left( {{x}_{i}} \right),\frac{i}{n} \right)-\left( {{z}^{0}}\left( {{x}_{i}} \right)+\frac{i}{n}*{{z}^{1}}\left( {{x}_{i}} \right) \right) \right)}}^{2}}
\end{equation}
where $n$ denotes the number of units, and we use ${}^{i}/{}_{n}$ to replace $t_{i}$ in the process of calculating MSE by simulating Monte Carlo sampling.

    \begin{table}[t]
    \centering
    \caption{Hyper-parameters Space of DeR-CFR}
        \begin{tabular}{c c}
            \hline
            {\textbf{Hyper-parameters}} & \textbf{Values} \\
            \hline
            \hline
            the number of & \multirow{2}{*}{\{2, all\}} \\ the constrained layers $l$  \\
            \hline
            {batch norm} & \multirow{2}{*}{\{False, True\}} \\
            {rep normalization} \\
            \hline
            depth of layers of & \multirow{2}{*}{\{1, 2, 3, 5, 7\}} \\ \{$d_R$, $d_y$, $d_t$\} \\
            \hline
            hidden state dimension of & \multirow{2}{*}{\{32, 64, 128, 256\}} \\ \{$h_R$, $h_y$, $h_t$\} \\
            \hline
            {\{$\alpha$, $\beta$, $\gamma$, $\mu$, $\lambda$\}} & \{1e-3, 1e-2, 1e-1 1, 5, 10, 100\} \\
            \hline
        \end{tabular}%
    \label{tab:Hyper}%
    \end{table}%
    
    \begin{table*}[htbp]
    \centering
    \caption{Optimal Hyper-parameters}
        \begin{tabular}{ p{9.11em} c c c c c}
            \hline
             {\textbf{Hyper-parameters}} &  {\textbf{IHDP}} &  {\textbf{Jobs}} &  {\textbf{Twins}} &  {\textbf{Binary}} &  {\textbf{Continuous}} \\
            \hline
            \hline
             {$l$} &  {2} &  {2} & {all} &  {all} & {all} \\
             {batch norm} &  {False} &  {True} &  {True} &  {False} &  {False} \\
             {rep normalization} &  {True} &  {True} &  {True} &  {False}  &  {False} \\
            \hline
             {\{$d_R$, $d_y$, $d_t$\}} &  {[7, 4, 1]} &  {[5, 4, 1]} &  {[7, 7, 3]} &  {[2, 5, 5]} &  {[5, 2, 3]} \\
             {\{$h_R$, $h_y$, $h_t$\}} &  {[32, 256, 256]} &  {[32, 128, 128]} &  {[64, 64, 64]} &  {[256, 128, 128]} &  {[256, 64, 64]} \\
             {\{$\alpha$, $\beta$, $\gamma$, $\mu$, $\lambda$\}}  & [5, 100, 1, 10, 1e-2] & [1e-2, 1, 1e-2, 5, 1e-3] & [1e-2, 1e-3, 1e-3, 5, 5] & [1e-1, 1, 1, 10, 1] & [1e-1, 1, 1e-1, 1, 10] \\
            \hline
        \end{tabular}\\
        
        \label{tab:optimal}%
    \end{table*}%
    
\section{Pseudo-Code of DeR-CFR}
\label{section:pseudo}

    As mentioned in the DeR-CFR Algorithm, the overall architecture of the model consists of the following components:
    \begin{itemize}
      \item Three decomposed representation networks for learning latent factors, one for each underlying factor: $I(X)$, $C(X)$ and $A(X)$.
      \item Three decomposition and balancing regularizers for confounder identification and balancing: the first is for decomposing $A$ from $X$ with considering $A(X) \perp T$ and $A(X)$ should predict $Y$ as precisely as possible; the second is for decomposing $I$ from $X$ via constraining $I(X) \perp Y \mid T$, and $I(X)$ should be predictive to $T$; the last is designed for simultaneously balancing confounder $C(X)$ in different treatment arms.
      \item Two regression networks for potential outcome prediction, one for each treatment arm: ${h^{0}}(C(X),A(X))$ and ${h^{1}}(C(X),A(X))$.
    \end{itemize}
    
    We adopt an alternating training strategy to iteratively optimize the representation for confounder identification and sample weight for confounder balancing as:
    \begin{eqnarray}
    \mathcal{L_{-\omega}} \!\!\! &=& \!\!\! \mathcal{L}_R + \alpha \cdot \mathcal{L}_{A}
        + \beta \cdot \mathcal{L}_{I}
        + \mu \cdot \mathcal{L}_{O}
        + \lambda \cdot Reg\\
    \mathcal{L_{\omega}} \!\!\! &=& \!\!\! \mathcal{L}_R
        + \gamma \cdot \mathcal{L}_{C\_B}
        + \lambda \cdot Reg
    \end{eqnarray}

    We minimize $\mathcal{L_{-\omega}}$ using stochastic gradient descent to update the parameters of the representation and hypothesis network, and minimize $\mathcal{L_{\omega}}$  to update $\omega$. Algorithm \ref{algorithm} shows the details of the pseudo-code of DeR-CFR \footnote{The code is available at the anonymous link: \url{https://www.dropbox.com/sh/5m40z2vmthx0y10/AACXJFuOvgB24av1VqkrkmKRa?dl=0}}. 
    
    \begin{algorithm}[H]
	\caption{Decomposed Representations for CounterFactual Regression}
	\begin{algorithmic}[1]
	\label{algorithm}
		\STATE \textbf{Input:} Observational data $\left\{x_{i}, t_{i}, y_{i}^{F}\right\}_{i-1}^{N}$
		\STATE \textbf{Output:} $\hat{y}_0, \hat{y}_1$
		\STATE \textbf{Loss function:} $\mathcal{L}_{-\omega}$ and $\mathcal{L}_{\omega}$
		\STATE \textbf{Components:} Three representation learning networks \{$I, C, A$\}, two regression networks $h^0$ and $h^1$ for the potential outcomes, two network $g_I, g_A$ to enforce $I, A$ to predict Treatment and Factual outcome as precisely as possible.
		\FOR{$i=0, 1, 2, ...$}
		\STATE $\left\{x_{i}, t_{i}, y_{i}^{F}\right\}_{i=1}^{N} \rightarrow \{I(X), C(X), A(X)\}$
		\STATE $\{I(X)\} \rightarrow g_I(I(X)) \rightarrow \hat{t}$
		\STATE $\{A(X)\} \rightarrow g_A(A(X)) \rightarrow \hat{y}$
		\STATE $ {h^0}(C(X),A(X)), {h^1}(C(X),A(X)) \rightarrow \hat{y}^0, \hat{y}^1$
		\STATE update $\mathcal{W} \leftarrow {\rm{Adam}}\{\mathcal{L}_{-\omega}\}$
		\STATE update $\omega \leftarrow {\rm{Adam}}\{\mathcal{L}_{\omega}\}$
		\ENDFOR
	\end{algorithmic}
    \end{algorithm}
    \noindent where $\mathcal{W}$ is the the trainable parameter of \{$I, C, A$, $h^0$, $h^1$, $g_I$, $g_A$\}, $\omega$ is the trainable sample weights, and the maximum number of iterations is $\mathcal{I}=3000$.
    
    
    Hardware used: Ubuntu 16.04.5 LTS operating system with 2 * Intel Xeon E5-2678 v3 CPU, 384GB of RAM, and 4 * GeForce GTX 1080Ti GPU with 44GB of VRAM.
    
    Software used: Python with TensorFlow 1.15.0, NumPy 1.17.4, and MatplotLib 3.1.1.

\section{Detailed Description of Real-world Data}
\label{section:real-word}
    \subsection{Semi-synthetic Benchmark: IHDP}
    The original Randomized Controlled Trial (RCT) data of the Infant Health and Development Program (\textbf{IHDP} \footnote{\url{http://www.fredjo.com/}} ) aims at evaluating the effect of specialist home visits on the future cognitive test scores of premature infants. Hill \cite{hill2011bayesian} removed a non-random subset of the treated group and induced selection bias. The dataset comprises 747 units (139 treated, 608 control) with 25 pre-treatment variables related to the children and their mothers. We report the estimation errors on the same benchmark (100 realizations of the outcomes with 63/27/10 proportion of train/validation/test splits) provided by and used in \cite{johansson2016learning,shalit2017estimating,hassanpour2020learning}. 
    
    \subsection{Real-world Data: Jobs}
    The \textbf{Jobs} \footnote{\url{http://www.fredjo.com/}} dataset created by LaLonde \cite{lalonde1986evaluating} is a widely used benchmark in the causal inference community, based on the randomized controlled trials. The dataset aims to estimate the effect of job training programs on employment status. Jobs contains 17 variables, such as age, education level, etc. Following Smith and Todd \cite{smith2005does}, we use LaLonde’s data (297 treated, 425 control) and the PSID comparison group (2490 control) to carry out our experiment. We randomly split the data of 3212 samples into train/validation/test with a 56/24/20 ratio (10 realizations). 
    
    \subsection{Real-world Data: Twins}
    The original \textbf{Twins} \footnote{\url{http://www.nber.org/data/linked-birth-infant-death-data-vital-statistics-data.html}} dataset is derived from the all twins born in the USA between the year of 1989 and 1991 \cite{almond2005costs}.
    
\section{Hyper-parameter Optimization}
\label{section:hyper}
    This algorithm selects ELU as the non-linear activation function and adopts Adam optimizer to minimize DeR-CFR’s objective function with a learning rate of 1e-3. We assign an adaptive weight to each unit in the training process and regard all samples as one full-batch. The maximum number of iterations is 3000. Table \ref{tab:Hyper} states the number and range of values tried per hyper-parameter during the paper's development. We return the best-evaluated iterate with early stopping and optimize the hyper-parameters in DeR-CFR by minimizing objective loss.
    
    Bergstra et al. \cite{bergstra2012random} demonstrated that trials on random search would be more efficient than grid search for optimizing hyper-parameter. In this paper, we randomly choose trails to determine the best Hyper-parameters for each Dataset within the Hyper-parameters space (Tabel \ref{tab:Hyper}). In addition, we will prioritize to fix model capacity [$d_R$, $d_y$, $d_t$, $h_R$, $h_y$, $h_t$] and select norm operations based on $\alpha=\beta=\gamma=\mu=\lambda=0, k=$all. And then, we proceed to the other Hyper-parameters search to optimize our model. 
    Tabel \ref{tab:optimal} lists all optimal hyper-parameters of DeR-CFR used for each dataset in the paper’s experiments.









\end{document}